\documentclass[apj]{emulateapj}

\newcommand\degree{\degr}
\newcommand\degrees\degree

\DeclareSymbolFont{UPM}{U}{eur}{m}{n}
\DeclareMathSymbol{\umu}{0}{UPM}{"16}
\let\oldumu=\umu
\renewcommand\umu{\ifmmode\oldumu\else\math{\oldumu}\fi}
\newcommand\micro{\umu}
\renewcommand\micron{\micro m}
\newcommand\microns \micron

\renewcommand\arcsec[0]{$^{\prime\prime}$}

\let\oldsim=\sim
\renewcommand\sim{\ifmmode\oldsim\else\math{\oldsim}\fi}
\let\oldpm=\pm
\renewcommand\pm{\ifmmode\oldpm\else\math{\oldpm}\fi}
\newcommand\by{\ifmmode\times\else\math{\times}\fi}

\newbox{\wdbox}
\renewcommand\c{\setbox\wdbox=\hbox{,}\hspace{\wd\wdbox}}
\renewcommand\i{\setbox\wdbox=\hbox{i}\hspace{\wd\wdbox}}

\newcount\timect
\newcount\hourct
\newcount\minct
\newcommand\now{\timect=\time \divide\timect by 60
         \hourct=\timect \multiply\hourct by 60
         \minct=\time \advance\minct by -\hourct
         \number\timect:\ifnum \minct < 10 0\fi\number\minct}

\newcommand\mctc{\multicolumn{2}{c}}

\catcode`@=11

\newcommand\comment[1]{}
\newcommand\commenton{\catcode`\%=14}
\newcommand\commentoff{\catcode`\%=12}

\renewcommand\math[1]{$#1$}
\newcommand\mathshifton{\catcode`\$=3}
\newcommand\mathshiftoff{\catcode`\$=12}

\comment{the backslash is necessary}

\comment{alignment tab}

\let\atab=&
\newcommand\atabon{\catcode`\&=4}
\newcommand\ataboff{\catcode`\&=12}

\let\oldmsp=\sp
\let\oldmsb=\sb
\def\sp#1{\ifmmode
           \oldmsp{#1}%
         \else\strut\raise.85ex\hbox{\scriptsize #1}\fi}
\def\sb#1{\ifmmode
           \oldmsb{#1}%
         \else\strut\raise-.54ex\hbox{\scriptsize #1}\fi}
\newbox\@sp
\newbox\@sb
\def\sbp#1#2{\ifmmode%
           \oldmsb{#1}\oldmsp{#2}%
         \else
           \setbox\@sb=\hbox{\sb{#1}}%
           \setbox\@sp=\hbox{\sp{#2}}%
           \rlap{\copy\@sb}\copy\@sp
           \ifdim \wd\@sb >\wd\@sp
             \hskip -\wd\@sp \hskip \wd\@sb
           \fi
        \fi}
\def\msp#1{\ifmmode
           \oldmsp{#1}
         \else \math{\oldmsp{#1}}\fi}
\def\msb#1{\ifmmode
           \oldmsb{#1}
         \else \math{\oldmsb{#1}}\fi}
\def\supon{\catcode`\^=7}
\def\supoff{\catcode`\^=12}
\def\subon{\catcode`\_=8}
\def\suboff{\catcode`\_=12}
\def\supsubon{\supon \subon}
\def\supsuboff{\supoff \suboff}

\newcommand\actcharon{\catcode`\~=13}
\newcommand\actcharoff{\catcode`\~=12}

\newcommand\paramon{\catcode`\#=6}
\newcommand\paramoff{\catcode`\#=12}

\comment{And now to turn us totally on and off...}

\newcommand\reservedcharson{\commenton \mathshifton \atabon \supsubon \actcharon
	\paramon}

\newcommand\reservedcharsoff{\commentoff \mathshiftoff \ataboff
	\supsuboff \actcharoff \paramoff}

\catcode`@=12
\reservedcharsoff

\reservedcharson

\comment{ Must have ONLY ONE of these... trust these macros, they work

}

\newcommand{\squishlist}{
 \begin{list}{$\bullet$}
  { \setlength{\itemsep}{1pt}
     \setlength{\parsep}{0pt}
     \setlength{\topsep}{3pt}
     \setlength{\partopsep}{0pt}
     \setlength{\leftmargin}{2.0em}
     \setlength{\labelwidth}{1.5em}
     \setlength{\labelsep}{0.5em} } }

\newcommand{\squishend}{
  \end{list}  }

\reservedcharson
\actcharon

\shorttitle{The WFC3 Emission Spectrum of Hot Jupiter HD209458b}
\shortauthors{Line {\em et al.}}

\begin{document}

\title{No Thermal Inversion and a Solar Water Abundance for the Hot Jupiter HD209458b from \emph{HST}/WFC3 Spectroscopy}


\author{\textsc{Michael R. Line\altaffilmark{1,2,3,4}, Kevin B. Stevenson\altaffilmark{5,6}, Jacob Bean\altaffilmark{6}, Jean-Michel Desert\altaffilmark{7}, Jonathan J. Fortney\altaffilmark{8}, Laura Kreidberg\altaffilmark{6}, Nikku Madhusudhan\altaffilmark{9}, Adam P. Showman\altaffilmark{10}, Hannah Diamond-Lowe\altaffilmark{11} }}
\altaffiltext{1}{NASA Ames Research Center, Moffet Field, CA 94035, USA}
\altaffiltext{2}{Bay Area Environmental Research Institute, 625 2nd Street Ste. 209, Petaluma, CA 94952, USA}
\altaffiltext{3}{Hubble Postdoctoral Fellow}
\altaffiltext{4}{School of Earth \& Space Exploration, Arizona State University}
\altaffiltext{5}{Sagan Postdoctoral Fellow}
\altaffiltext{6}{Department of Astronomy \& Astrophysics, University of Chicago, 5640 S Ellis Ave, Chicago, IL 60637, USA}
\altaffiltext{7}{University of Amsterdam}
\altaffiltext{8}{Department of Astronomy \& Astrophysics, University of California, Santa Cruz, 1156 High Street, Santa Cruz, CA 95064, USA}
\altaffiltext{9}{Institute of Astronomy, University of Cambridge, Cambridge CB3 0HA, UK}
\altaffiltext{10}{Department of Planetary Sciences and Lunar and Planetary Laboratory, University of Arizona, 1629 E. University Blvd., Tucson, AZ 85721, USA}
\altaffiltext{11}{Department of Astronomy,Harvard Smithsonian Center for Astrophysics 60 Garden Street, MS-10 Cambridge, MA 02138, USA  }

\begin{abstract}
The nature of the temperature-pressure profiles of hot Jupiter atmospheres
is one of the key questions raised by the characterization of transiting
exoplanets over the last decade. There have been claims that many
hot Jupiters exhibit atmospheric thermal inversions where there is an
increasing temperature with decreasing pressure in the infrared
photosphere that leads to the reversal of molecular absorption bands into
emission features. However, these claims have been based on
broadband photometry rather than the unambiguous identification of
emission features with spectroscopy, and the chemical species that
could cause the thermal inversions by absorbing stellar irradiation at
high altitudes have not been identified despite extensive theoretical and
observational effort. Here we present ultra-high precision \emph{HST} WFC3 observations of the dayside thermal emission spectrum of the hot Jupiter HD209458b, which was the first exoplanet suggested to have
a thermal inversion. In contrast to previous results for this planet,
 our observations resolve a water band in absorption at 6.2$\sigma$ confidence. When combined with \emph{Spitzer} photometry the data are
indicative of a monotonically decreasing temperature with pressure
over the range 1-0.001 bar at 7.7$\sigma$ confidence.  We test the robustness of our results by exploring a variety of model assumptions including the temperature profile parameterization, presence of a cloud, and choice of \emph{Spitzer} data reduction. We also introduce a new analysis method, {\it chemical retrieval-on-retrieval}, to determine the elemental abundances from the spectrally retrieved mixing ratios with thermochemical self-consistency and find plausible abundances consistent with solar metallicity (0.06 - 10$\times$ solar) and carbon-to-oxygen ratios less than unity. This work suggests that high-precision spectrophotometric results are required to robustly infer thermal structures and compositions of extra-solar planet atmospheres, and to perform comparative exo-planetology. 

\end{abstract}

\section{Introduction}\label{sec:intro}

\emph{Hubble Space Telescope} (\emph{HST}) Wide Field Camera 3 (WFC3) spectrophotometric observations of transiting exoplanets are rapidly transforming our knowledge of planetary atmospheres (Berta et al. 2012; Swain et al. 2012; Deming et al. 2013; Huitson et al. 2013; Kreidberg et al. 2014ab,15; Stevenson et al. 2014c; McCullough et al. 2014; Knutson et al. 2014; Fraine et al. 2014; Evans et al. 2016). As the highest precision and most robust spectroscopy obtainable with current instruments, WFC3 data allow us to infer the thermal structures, energy transport, molecular abundances, and cloud properties in exoplanet atmospheres. Determining these properties allows us to then begin to understand the fundamental processes occurring in planetary atmospheres over a wide range of conditions. 

The thermal structures of hot Jupiter atmospheres, and in particular the question of thermal inversions has driven much of the characterization work on these planets over the last decade  (for reviews, see Burrows \& Orton, 2010; Madhusudhan et al. 2014a; Crossfield et al. 2015).
Thermal inversions typically occur in the presence of a high altitude heating source, usually a strong shortwave absorber.   Most solar system bodies with substantial atmospheres possess a thermal inversion (e.g., Robinson \& Catling 2014), due to ozone in the case of Earth, or aerosol heating in the case of the giant planets.   Hot Jupiters were theorized to have thermal inversions due to the possible presence of a strong gas phase optical absorbers (Hubeny et al. 2003; Burrows et al. 2008 ), one example being TiO and VO, which can persist at high temperatures (T$>$2000 K at 1 bar, Fortney et al. 2006, Fortney et al. 2008 ), or poly-sulfur aerosols (Zahnle et al. 2009).

Observations using 6 broadband channels on the \emph{Spitzer} Space Telescope claimed the first detection of a thermal inversion in a hot Jupiter atmosphere, in the prototype first transiting planet HD 209458b, (Knutson et al. 2008, Burrows et al.\ 2007), presumably confirming the presence of a high-altitude absorber, the leading candidates being TiO \& VO their presence in the gas phase at HD209458b's temperatures.  Evidence for the inversion was based upon the relatively high fluxes of the 4.5 and 5.8 $\mu$m channels compared with the 3.6 and 8.0 $\mu$m channels. These results were challenging to interpret theoretically as, first, for plausible compositions, the contribution functions of the four Spitzer channels greatly overlap, so it is extremely difficult to get significantly differing brightness temperatures between the channels, and secondly,  the infrared photosphere (optical depth of unity) tends to occur at deeper layers than the layers at which most of the visible layers is absorbed. An inversion is formed because at the altitudes at which the visible radiation is strongly absorbed the emissivity is low, requiring hotter temperatures to maintain radiative equilibrium. Never-the-less, Madhusudhan \& Seager (2009) and Line et al. (2014), using retrieval methods, were able to fit the Knutson et al. (2008) Spitzer data by invoking an inversion above the $\sim$1 bar level and large abundances of CO. The large abundance of CO is required to decouple the 4.5{\micron} (and to a lesser extent 5.8 $\micron$) weighting functions from the rest of the channels, permitting the large temperature differences.  It is unclear, however, how well the mean infrared opacity required to produce the retrieved thermal structures agrees with those derived from the abundances.

Diamond-Lowe et al. (2014) recently  re-evaluated the thermal inversion hypothesis for HD209458b. They analyzed the Knutson et al. (2008) dataset with new data reduction methods, and also analyzed new data taken with more optimal observing modes at 3.6, 4.5, and 8.0 $\mu$m. This new study resulted in a significant reduction of the 4.5 and 5.8 $\mu$m fluxes, and showed that a thermal inversion was no longer required to model those data. Additionally, Schwarz et al. (2015), using hi-dispersion spectroscopy with the CRIRES instrument combined with cross-correlation data reduction techniques, were able to rule out an inversion between 1 bar and 1 mbar. However, they also did no€™t make a strong detection of molecular absorption indicative of a normal thermal structure.

Furthermore, evidence for inversions in other planets of similar or hotter temperatures is currently lacking.  In the years since the HD 209458b observations, retrieval methods of interpreting spectra have shown that often, models without temperature inversions could explain wide-band \emph{Spitzer} data as readily as models with inversions (Madhusudhan \& Seager 2009; Line et al. 2014).  A growing body of work has tried to explain why inversions may not occur.  Spiegel et al. (2009) and Showman et al. (2009) investigated how condensation in cold traps could lead to removal of TiO and VO from the gas phase. Madhusudhan et al. (2011) suggested the possible lack of inversions to be due to high carbon-to-oxygen (C/O) ratios. High C/O would deplete the oxygen required to build the strongly absorbing oxides.  However, Line et al. (2014) and Benneke (2015) showed that there is no strong evidence for high C/O ratios over a wide range of planetary conditions--including those in which inversions are expected.  An alternative hypothesis for the apparent lack of inversions, brought forth by Knutson et al. (2010), suggests that high incident UV flux could potentially photolize high altitude absorbers. 

Recently, Haynes et al. (2015) suggested the presence of an inversion along with potential evidence for TiO emission in the Hot-Jupiter WASP-33b by combining \emph{HST}, \emph{Spitzer}, and ground based data,   While tantalizing evidence, a bulk of the inversion evidence is driven by the ground based z' band point and the two bluest channels in WFC3.  Furthermore, WASP-33 is a delta-scuti variable, thus making the data analysis and interpretation extremely challenging.

The power of spectroscopy, compared to photometry, is that the planetary thermal emission is probed across a range of wavelengths, and hence a range of depths in the atmosphere, simultaneously.  The interpretation of such observations is in principle much more straightforward than in wide-band photometry.  Molecular features seen in emission can be interpreted as a tell-tale sign of an inverted temperature structure.  However, a high S/N spectrum is required.  As part of \emph{HST} GO program 13467 we have endeavored to produce such high S/N spectral data to bring clarity to the characterization of exoplanet atmospheres. 

 In what follows, we describe our WFC3 observations in \S\ref{sec:obs}, followed by a global dayside atmospheric retrieval analysis \S\ref{sec:retrieve}, and finally we discuss implications of our observations and conclude in \S\ref{sec:conc}

\section{\emph{HST}/WFC3 OBSERVATIONS AND DATA ANALYSIS}\label{sec:obs}

\subsection{Observations and Reduction}


As part of a large {\em HST} Treasury Program (GO 13467), we used the Wide Field Camera 3 (WFC3) to observe HD~209458b during secondary eclipse over five visits from September to December of 2014.  WFC3's G141 grism disperses light from 1.12 -- 1.65 $\mu$m onto its detector, which we operated in 256$\times$256 subarray mode.

Each visit comprises five {\em HST} orbits: the first orbit achieves instrument stability and is excluded from our analysis following standard procedure (Berta et al. 2012, Deming et al. 2013, Kreidberg et al. 2014b), the second and fifth orbits determine the out-of-eclipse baseline, and the third and fourth orbits contain the secondary eclipse.  Each {\em HST} orbit begins with a direct image of HD~209458 for calibration purposes, is followed by 43 spectroscopic frames (each lasting 14.971 seconds in duration), and concludes with a buffer dump to maximize observing efficiency.  During the acquisition of each spectroscopic frame, we utilize the bidirectional spatial scan mode at a rate of 1.15~\arcsec/s, thus spreading the light over $\sim$140 pixel rows perpendicular to the spectral dispersion.  The median peak pixel count across all frames and all visits is 40,615e, which is close to the recommended threshold level of $\sim$40,000e for optimal performance (Berta et al. 2012, Swain et al. 2013, Wilkins et al. 2014).  

Our data reduction and analysis closely follows the techniques described in our previous WFC3 papers (Stevenson et al. 2014a,c,d).  Here we provide a brief description of our methods.  For the reduction of spatial scan data, we use the calibrated ``\_ima'' frames provided by the STScI Archive.  Using a 2D Gaussian, we compute the centroid of the direct image from the start of each {\em HST} orbit, calculate the trace of the equivalent stare spectra, and then determine the field-dependent wavelength solution of the scanned spectra (Kuntschner et al. 2009).  We apply spectroscopic flat field corrections using coefficients from the calibration file {\tt WFC3.IR.G141.flat.2.fits}.

Each frame consists of three evenly-spaced, non-destructive reads.  We compute the difference between pairs of adjacent reads, use optimal extraction (Horne 1986) to produce two 1D spectra, and then combine their values to arrive at a single spectrum per frame.  Our extraction window consists of 110 pixel rows centered on the differenced spectra and flanked by an additional 110 pixel rows for background subtraction (220 pixel rows total).  This optimized configuration produces the most precise light curves. We divide the spectra into 10 spectrophotometric bins of width $\sim$0.0525 $\mu$m and spanning 1.125 -- 1.655 $\mu$m.

Our extracted light curves exhibit similar systematics to those noted previously for WFC3 data (e.g., Berta et al. 2012, Deming et al. 2013, Stevenson et al. 2014c).  These include a ramp (or hook) at the beginning of each {\em HST} orbit and a visit-long trend.  We follow a two step process to extract HD~209458b's emission spectrum.  First, we model the band-integrated (white) light curves to compute non-analytic, transit-removed systematic model components and to evaluate the consistency of the measured eclipse depths between visits.  Second, we divide the spectroscopic light curves by our non-analytic model components and then determine the best fit for each channel to derive wavelength-dependent eclipse depths with uncertainties.  Below we discuss each of these steps in detail.

\subsection{White Light-Curve Fits}\label{sec:whiteFits}

Our model fitting procedure consists of finding the best solution by simultaneously fitting all free parameters using a Levenberg-Marquardt minimizer and then using a Differential-Evolution Markov Chain algorithm (DEMC, ter Braak et al. 2008) to estimate parameter uncertainties.  We find no evidence for correlated noise in the best-fit residuals.  We test various model components and shared parameters then determine the best combination by comparing their Bayesian Information Criterion values (BIC, Cornish \& Littenberg 2007).

Our final white light curve model takes the form:
\begin{equation}
\label{eqn:full}
F(\Updownarrow,\phi,\Phi) = F\sb{\rm s}(\Updownarrow)E(\phi)R(\phi,\Updownarrow)H(\Phi,\Updownarrow),
\end{equation}
\noindent where $F(\Updownarrow,\phi,\Phi)$ is the modeled flux for scan direction, $\Updownarrow$, at planet orbital phase, $\phi$, and {\em HST} orbital phase, $\Phi$; $F\sb{\rm s}(\Updownarrow)$ is the scan-direction-dependent out-of-eclipse system flux; $E(\phi)$ is the eclipse model (Kreidberg 2015); $R(\phi,\Updownarrow) = 1 + r\sb{1,\Updownarrow}(\phi-0.5) + r\sb{2,\Updownarrow}(\phi-0.5)^{2}$ is the time-dependent quadratic trend model component; and $H(\Phi,\Updownarrow) = 1 - e\sp{-h\sb{0}\Phi + h\sb{1}} + h\sb{2,\Updownarrow}\Phi$ is the {\em HST} orbital phase-dependent rising exponential ramp with linear trend model component. Scan-direction- and visit-dependent free parameters include $F\sb{\rm s}(\Updownarrow)$, $r\sb{1,\Updownarrow}$, $r\sb{2,\Updownarrow}$ and $h\sb{2,\Updownarrow}$.  In our final fits, the eclipse depth, $h\sb{0}$, and $h\sb{1}$ free parameters have the same values for both scan directions and all visits.  We fix the eclipse midpoint ($\phi = 0.5$), total eclipse duration (3.06 hours) and ingress/egress times (0.42 hours, Seager et al. 2003) because these parameters are poorly constrained due to the long observing gaps during Earth occultation, but also precisely known from previous observations of this canonical planet (Zellem et al. 2014).

\begin{table}[tb]
\centering
\caption{\label{tab:ObsDates} 
White Light Curve Eclipse Depths}
\begin{tabular}{ccr@{\,{\pm}\,}lc}
    \hline
    \hline
    Visit \#    & Observation Date  & Eclipse Depth & $\chi^2_{\nu}$\\
                &                   & (ppm)          &               \\
    \hline
    1           & 2014 September 16 & \hspace{8pt}113 & 36  & 1.11          \\
    2           & 2014 October 18   &             121 & 35  & 0.87          \\
    3           & 2014 November 15  &              50 & 28  & 1.43          \\
    4           & 2014 December 10  &               6 & 35  & 1.18          \\
    5           & 2014 December 31  &             122 & 34  & 1.15          \\
    \hline
    (1,2,3,4,5) & --                &   82 & 15             & 1.14          \\
    (1,2,3,5)   & --                &   99 & 17\tablenotemark{a} & 1.13          \\
    \hline
\end{tabular}
\tablenotetext{1}{Adopted value.}
\end{table}

\begin{figure}[tb]
\centering
\includegraphics[width=1.0\linewidth,clip]{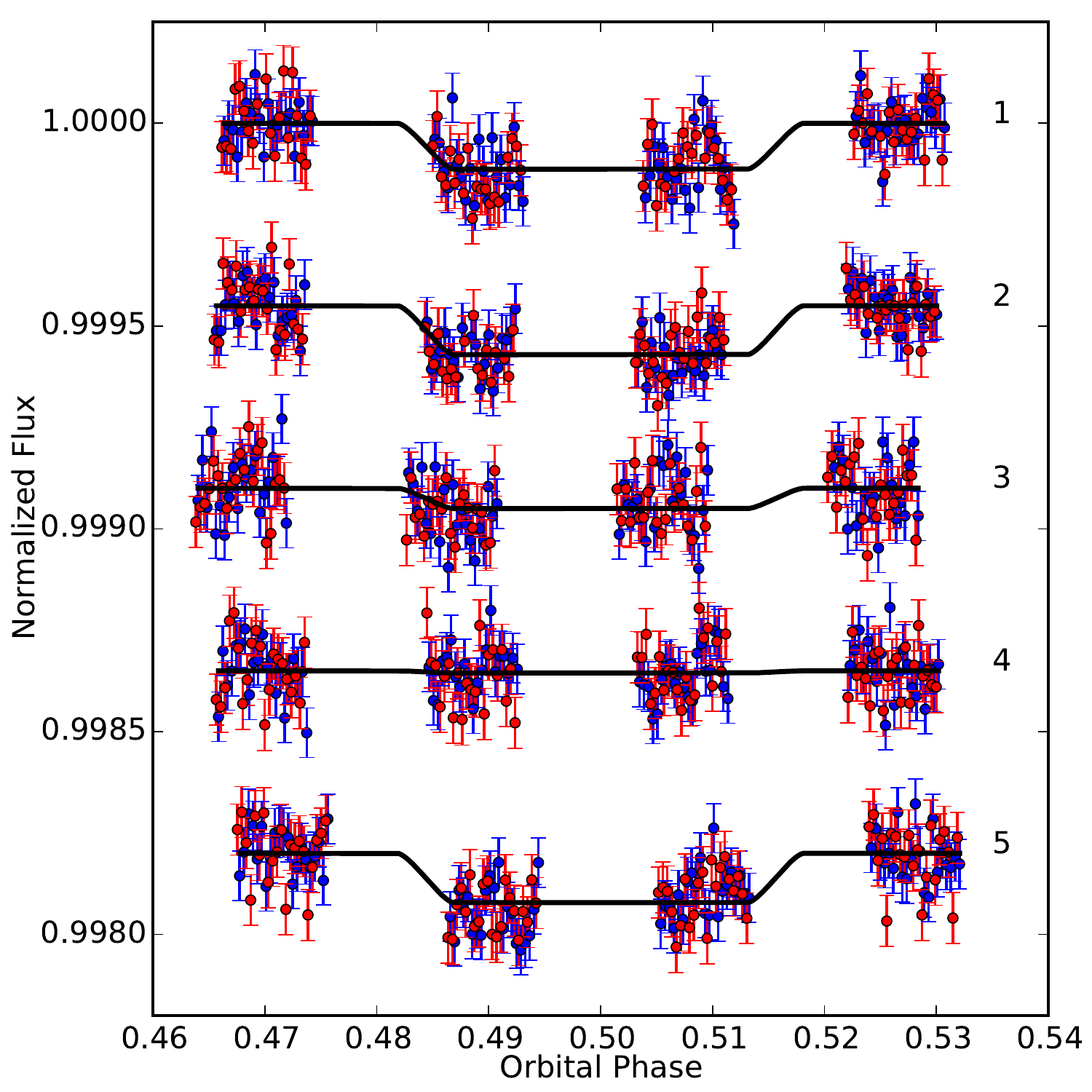}
\caption{\label{fig:LCs}{
White light curves of HD~209458b.  The blue and red data points depict the systematics-removed normalized flux values from the forward and reverse scan directions, respectively.  The best-fit models (black lines) highlight the indistinguishable eclipse depth during the fourth visit (labels indicate the visit \#).
}}
\end{figure}

To test the repeatability of our measurements, we apply our white light curve model to the data from each visit and report the best-fit eclipse depths and $\chi^2_{\nu}$ values in Table \ref{tab:ObsDates}.  Despite having a relatively good fit ($\chi^2_{\nu} = 1.18$) and contrary to the other visits, the fourth visit contains no discernible secondary eclipse (6 {\pm} 35~ppm, Figure \ref{fig:LCs}).  As noted previously (Stevenson et al. 2014c), there is a strong correlation between the eclipse depth and the quadratic term in $R(\phi,\Updownarrow)$.  As a test, we include data from the latter half of the first {\em HST} orbit and derive a deeper eclipse depth (54 ppm) and shallower $r\sb{2,\Updownarrow}$ values (the other free parameters are unaltered).  We conclude that this parameter degeneracy is the most likely source of the fourth visit's eclipse-depth inconsistency.  Many WFC3 primary transit analyses emphasize obtaining robust relative transit depths; however, secondary eclipse analyses rely on the absolute depth to determine the planet's temperature.  This places additional emphasis on correctly handling the white light curve systematics.

\begin{figure}[tb]
\centering
\includegraphics[width=1.0\linewidth,clip]{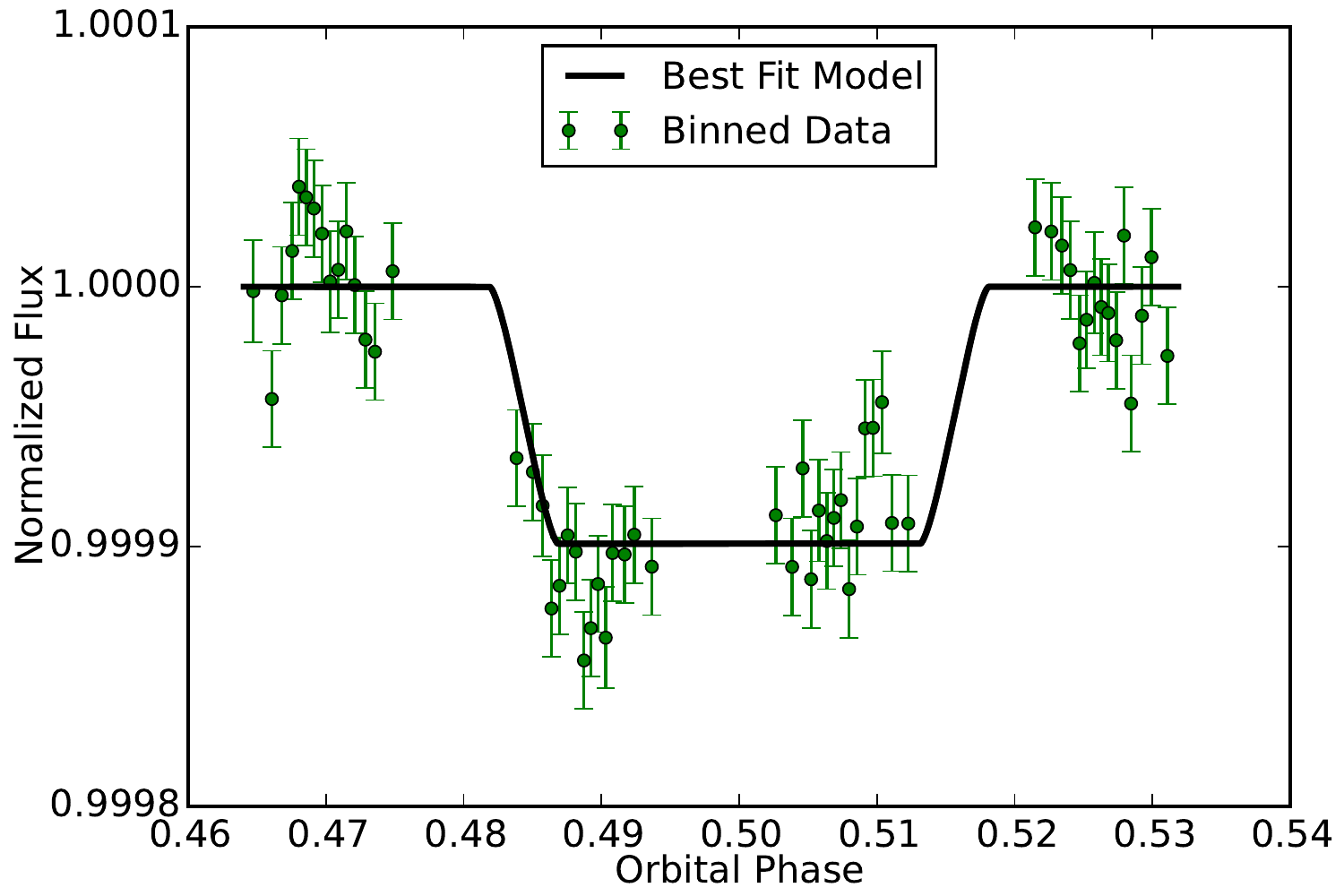}
\caption{\label{fig:JointLCs}{
Binned white light curve of HD~209458b.  The green points combine data from visits 1, 2, 3, and 5; the best-fit model is from our joint fit.
}}
\end{figure}

When we exclude the first {\em HST} orbit and include all of the visits in a joint fit, the best-fit model favors a shared eclipse depth of 82 {\pm} 15~ppm.  Adding data from the latter half of the first {\em HST} orbit to all visits increases the shared eclipse depth to 97 {\pm} 15~ppm.  This value is comparable to a fit that excludes the first {\em HST} orbit and the fourth visit (99 {\pm} 17~ppm).  We choose to adopt this final value as our estimate of the true white light curve eclipse depth (Table \ref{tab:ObsDates}).  Figure \ref{fig:JointLCs} depicts systematics-removed, binned data (excluding visit 4) with a best-fit light curve model from our joint fit.

\subsection{Spectroscopic Light-Curve Fits}
\label{sec:specFits}

We apply the {\tt Divide-White} technique (Stevenson et al. 2014a) to remove the wavelength-independent systematics from the spectroscopic light curves.  This is accomplished by dividing the spectroscopic light curves from each visit by their corresponding, transit-removed white light curves, which serve as non-analytic models of the wavelength-independent systematics.
Our final spectroscopic light curve model takes the form:
\begin{equation}
\label{eqn:spec}
F(\lambda,\Updownarrow,\phi) = F\sb{\rm s}(\lambda,\Updownarrow)E(\lambda,\phi)R(\lambda,\phi)W(\phi),
\end{equation}
where $F(\lambda,\Updownarrow,\phi)$ (the wavelength-dependent modeled flux), no longer depends on {\em HST}'s orbital phase, the definitions for $F\sb{\rm s}(\lambda,\Updownarrow)$ and $E(\lambda,\phi)$ are unchanged, $R(\lambda,\phi) = 1 + r\sb{\lambda,1}(\phi-0.5)$ is now a time-dependent linear trend with no scan-direction dependence on $r\sb{\lambda,1}$, and $W(\phi)$ is the non-analytic, transit-removed white light curve model.  The simplicity of this model relative to Equation~\ref{eqn:full} highlights the effectiveness of using the {\tt Divide-White} technique to analyze {\em HST}/WFC3 data.

In Table \ref{tab:Depths}, we provide the spectroscopic eclipse depths from a joint fit encompassing visits 1, 2, 3, and 5.  Adding the fourth visit effectively reduces the eclipse depths by an average of $\sim$17~ppm.  The listed eclipse depth uncertainties are relative values; therefore, a wavelength-independent offset should be applied when fitting atmospheric models that consider additional data.  In this case, a Gaussian prior of width 17~ppm (the white light curve eclipse depth uncertainty) is appropriate.  Table \ref{tab:Depths} also lists mean $\chi^2_{\nu}$ values for each channel; the overall mean is 1.04 times the photon plus read-noise limit.  In the next section we interpret these measurements within an atmospheric retrieval framework.

\begin{table}[tb]
\centering
\caption{\label{tab:Depths} 
Spectroscopic Light Curve Eclipse Depths}
\begin{tabular}{cr@{\,{\pm}\,}lc}
    \hline
    \hline
    Wavelength      & \mctc{Eclipse Depth}  & $\chi^2_{\nu}$\\   
    ($\mu$m)    & \mctc{(ppm)}          &               \\
    \hline
    1.125 -- 1.178  & \hspace{11pt}43 & 16  & 0.98          \\
    1.178 -- 1.230  &              97 & 15  & 1.01          \\
    1.230 -- 1.282  &             115 & 15  & 1.02          \\
    1.282 -- 1.335  &             109 & 15  & 1.13          \\
    1.335 -- 1.388  &              80 & 15  & 1.00          \\
    1.388 -- 1.440  &              43 & 15  & 0.96          \\
    1.440 -- 1.492  &              86 & 15  & 0.84          \\
    1.492 -- 1.545  &             107 & 17  & 1.11          \\
    1.545 -- 1.598  &             167 & 17  & 1.20          \\
    1.598 -- 1.650  &             170 & 18  & 1.12          \\
    \hline
\end{tabular}
\end{table}


\section{Atmospheric Retrieval Analysis}\label{sec:retrieve}
Our new WFC3 measurements combined with \emph{Spitzer} broadband photometry previously presented by our group (Diamond-Lowe et al. 2014) are interpreted within the context of an atmospheric retrieval framework. In what follows we first describe our nominal model setup (\S\ref{sec:setup}) and results (\S\ref{sec:results}). We also test the robustness of our nominal model assumptions by exploring a different temperature-profile parameterization (\S\ref{sec:TP_par}), the inclusion of a dayside cloud (\S\ref{sec:cloud}), and the impact of different \emph{Spitzer} data reduction methods (\S\ref{sec:spitzer}).  Finally, we explore the chemical plausibility of our retrieved abundances through a new method, { \it chemical retrieval-on-retrieval} (\S\ref{sec:ronr}).

\subsection{Atmospheric Parameterization}\label{sec:setup}
We use a derivative of the emission portion of the CHIMERA retrieval suite (Line et al. 2013; 2014; Stevenson et al. 2014c; Kreidberg et al. 2014b; Diamond-Lowe et al. 2014) to interpret the full dayside emission data of HD~209458b.  The CHIMERA emission forward model computes the upwelling disk-integrated thermal emission given a temperature structure and molecular abundances and has been validated against other models in two publications (Line et al. 2013; Morley et al. 2015).  This forward model is coupled with the powerful {\tt pymultinest} routine (Buchner et al. 2014), a {\tt python} wrapper to the commonly used {\tt multinest} nested sampling algorithm  (Skilling 2006; Feroz et al. 2009), to perform the parameter estimation and Bayesian evidence computation. 

The temperature-pressure (T-P) profile is parameterized (with 5 parameters) using the analytic approximation from Parmentier \& Guillot 2014 (see Line et al. 2013 for implementation details).  The parameterization is flexible enough to permit a wide range of thermal structures, including inversions, while also resulting in physically realistic profiles consistent with radiative energy balance as the analytic approximation assumes radiative equilibrium in its derivation.  For this investigation we retrieve 7 thermochemically plausible gases that absorb over the wavelengths of interest: H$_2$O, CH$_4$, CO, CO$_2$, NH$_3$, HCN, \& C$_2$H$_2$ (e.g., Madhusudhan et al. 2011) with uniform-in-log mixing ratio priors spanning from -12 to 0. Molecular hydrogen and helium (in solar proportions) are assumed to comprise the remaining gas such that all species sum to unity.   We use the absorption cross-section database described in Freedman et al. (2014). For this analysis we make no {\it a priori} assumption about the realism of our retrieved abundances, but will later show that they are consistent with expectations from thermochemical equilibrium suggesting that both the data and model reasonably reflect reality. All emission model results presented in this work assume cloud free atmospheres unless otherwise stated. We also assume a free parameter offset between the WFC3 and \emph{Spitzer} data to account for uncertainties in the absolute calibration in the white light transit depth , though the impact of this parameter is minimal in our retrievals (found to be consistent with no offset, 3$\pm10$ppm).

\subsection{Fiducial Results}\label{sec:results}
\begin{figure*}[h]
\centering
\includegraphics[width=1.0\linewidth]{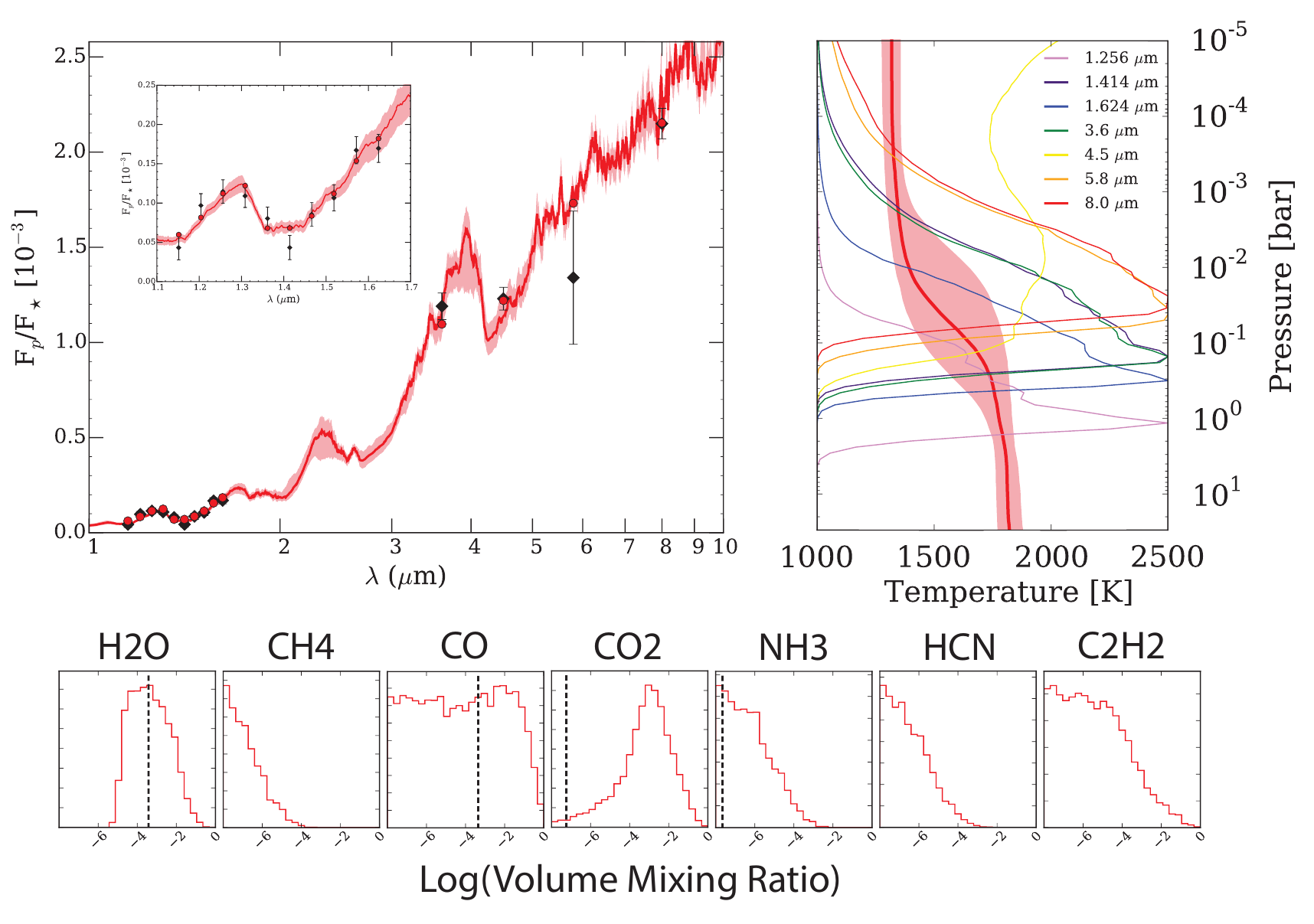}
\caption{\label{fig:spectra_TP}{
Nominal spectral fits, temperature profile, and molecular abundance retrieval results from the ``FULL" model.  Left: Model fits to both the WFC3 and \emph{Spitzer} data.The data (WFC3, this work, and \emph{Spitzer} from Diamond-Lowe et al. 2014) are shown as black diamonds with error bars.   The light red shaded regions are the 1-sigma spread in the spectra resulting from 1000 draws from the posterior and solid red curve is the median of those 1000 spectra. The red points are the medians of the  model spectra integrated over the photometric bandpasses (median $\chi^{2}$ per data point of 1.03) .  A zoom in over the WFC3 region is shown in the inset. Right:   Range of allowed TP profiles that best explain the data.    The light red shaded regions are the 1-sigma spread in the TP-profiles resulting from 1000 draws from the posterior and solid red curve is the median of those 1000 TP-profiles. This is a family of solutions for which all of the TP profiles are mono-tonically decreasing (no evidence for inversion). Normalized thermal emission contribution functions (chapman functions) are shown for select wavelengths. Note how the contributions functions are concentrated between $\sim$1 bar - 1 mbar.  The TP profile outside of this range is largely driven by the parameterization and should not be taken as absolute truth.   Bottom: Marginalized posterior distributions for each of the gases included in the nominal model.  The vertical dashed line in each panel is the predicted thermochemical equilibrium mixing ratio at 1700K, 0.1 bar under solar elemental abundances. The solar composition values for CH$_4$, C$_2$H$_2$, and HCN are less than the plot range lower limit. We find strong water vapor absorption in the WFC3 spectrum, with an abundance consistent with solar composition (within $\pm$1 dex), and a monotonically decreasing temperature profile between 1 bar and 1 mbar.
}}
\end{figure*}

\begin{figure}[h]
\centering
\includegraphics[width=1.0\linewidth]{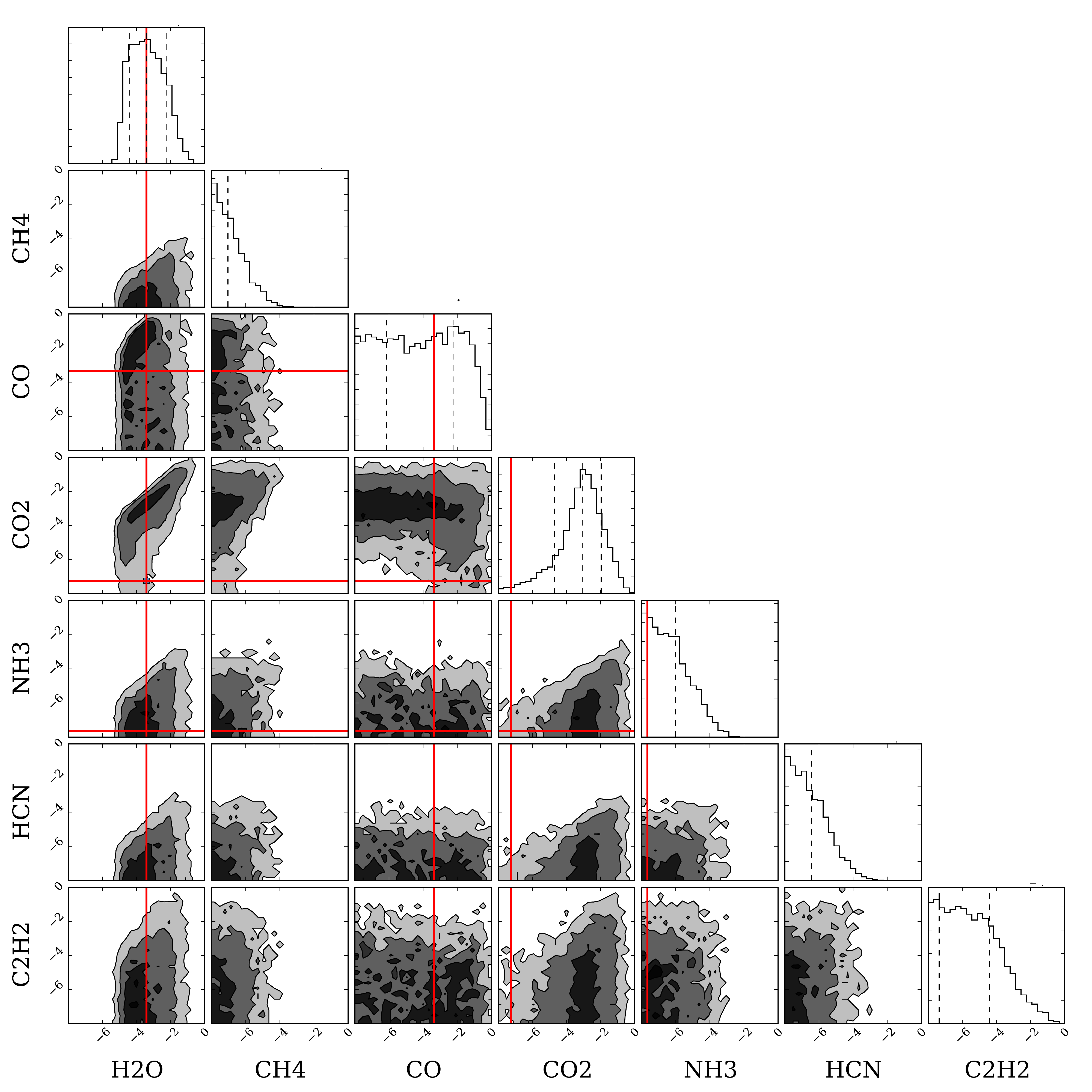}
\caption{\label{fig:stair_pairs}{
Nominal gas retrieval results from the ``FULL" model showing the correlations amongst each gas.  The black, dark gray, and light gray in the 2-D histograms correspond to the 1 (39.3\%)-, 2 (86.5\%)-, and 3 (98.9\%)-sigma confidence intervals.  The dashed vertical lines in the histograms are the marginalized 16, 50, and 84 percentiles. The vertical and horizontal red lines in each panel are the solar composition molecular abundances at 1700K and 0.1 bar, a representative photospheric temperature (see Figure \ref{fig:spectra_TP}) . The water mixing ratio is constrained to $\pm$1 order of magnitude. Note the ``elbow" shaped correlation between CO and CO$_2$. This is because CO and CO$_2$ both absorb over the 4.5 $\mu$m \emph{Spitzer} bandpass, and are thus highly degenerate.  
}}
\end{figure}

Figures \ref{fig:spectra_TP} and \ref{fig:stair_pairs} summarize our nominal retrieval results under the aforementioned model assumptions, denoted as the  {\it FULL} model scenario.  The spread in the models and fits suggests that the model and the data are well matched (median chi-squared per data point of 1.03) much like our previous results on WASP-43b, suggesting that our atmospheric parameterization is adequate, hence parameter estimates from this data-model combination have meaning.  It comes as no surprise that the presence of a deep water vapor absorption feature in the WFC3 bandpass results in a monotonically decreasing TP profile over the planet's infrared photosphere (1 bar - 1 mbar).  Water is the only robustly constrained molecular absorber, with an abundance constraint of $\pm$1 order of magnitude nearly centered about solar composition (68\% confidence interval: 4.0$\times 10^{-5}$ - 5.4$\times 10^{-3}$ ). We can only retrieve upper limits or long unconstrained tails (CO$_2$) on the other gases.  The resulting abundance distributions (Figure \ref{fig:stair_pairs} ) appear physically plausible (with the exception of CO$_2$, more below), or at least they encompass physically plausible molecular combinations, and are consistent with or near solar abundance to within 3$\sigma$. 

In order to further demonstrate the lack of need for a thermal inversion, and to identify which molecules are actually ``detected" in this spectrum, we perform a standard nested Bayesian model comparison using the full Bayesian evidence calculation from {\tt pymultinest} (Trotta 2008; Benneke \& Seager 2013; Swain, Line \& Deroo 2014; Waldmann et al. 2015a,b; Kreidberg et al. 2015).  The Bayesian evidence is what the commonly used Bayesian Information Criterion (BIC) approximates via a truncated Laplace approximation (Trotta 2008; Cornish \& Littenberg 2007).  A parameter is deemed ``necessary" or ``detected" if the improvement in likelihoods from the addition of that given parameter (or parameters) outweighs the increase in prior volume (Chapter 3.5, Gregory 2005).  This is a straightforward method for determining whether or not a gas is detected as we simply remove that gas from the model and re-run the retrieval to compute the new evidence.  The ratio of the evidences of two models (usually a {\it FULL} model with all parameters compared with subsets of that model) is the Bayes factor of which can be directly converted into a confidence interval/detection significance (equation 27, Table 2 in Trotta 2008; reproduced in Benneke \& Seager 2013).  

Table \ref{tab:Detections_Sigs} summarizes the detection significances of the various nested models. We find that water is detected at 6.2$\sigma$ confidence. This detection, as with our WASP-43b WFC3 observations (Kreidberg et al. 2014b), is strongly driven by the 1.4$\mu$m water feature covered by WFC3.   The reduced gases, expected to occur in cooler planets (CH$_4$, NH$_3$) or high carbon-to-oxygen ratio atmospheres (HCN, C$_2$H$_2$) are not detected, and in fact their inclusion is penalized due to the increase in prior volume without the accompanying improvement in fit quality, consistent with retrieving only upper limits on their abundances.  CO \& CO$_2$ are detected at 4.1 $\sigma$ confidence. This detection is largely driven by the 4.5 $\mu$m IRAC observation. We cannot decipher the difference between CO \& CO$_2$ as they have nearly overlapping features over the 4.5 $\mu$m bandpass, thus the removeal of one is compensated by the other.

To compute the ``detection"  (or lack there of) of a thermal inversion we perform a separate retrieval in which the priors on the 5-parameter analytic TP-profile force TP-profiles that have zero to positive temperature gradients with increasing altitude. In other words we restrict the prior search space to exclude monotonically decreasing profiles.  We have not removed or added parameters, we simply reduced the prior volume by restricting the prior ranges.  Upon comparing the Bayesian evidence of the two models, we find that a monotonically decreasing TP profile is required at the 7.7$\sigma$ level.

In summary, we have robustly detected absorption due to water at or near what is expected from solar elemental abundances ($\pm$1 dex), a detection of some combination of CO and CO$_2$, and strong evidence for a monotonically decreasing TP-profile over the pressure levels robustly\footnote{By robustly we mean where most of the integrated area of the weighting functions lie. There is some very weak contribution at lower pressure levels in the broad 4.5 $\mu$m IRAC band-pass.} probed by the observations (1bar - 1 mbar).  In the following sections we will explore how various model and data assumptions will impact these conclusions.

\begin{table}[h]
\centering
\caption{\label{tab:Detections_Sigs} 
Nested Bayesian model comparison results.  The scenarios indicate which molecules/setup we are detecting. Water, CO+CO$_2$, and a decreasing temperature profile are detected at high confidence. The full compliment of gases includes H$_2$O, CH$_4$, CO, CO$_2$, NH$_3$, HCN, C$_2$H$_2$}
\begin{tabular}{lccc}
    \hline
    \hline
    Scenario \tiny{(included gases)}     & $\ln B$  & Det. Sig. ($\sigma$)\\
    \hline
     FULL\tiny{(all gases)} & - & -\\
     Molecules other than Water \tiny{(H$_2$O)}   & 3.96 & 3.3 \\
     CO \& CO$_2$ \tiny{(H$_2$O, CH$_4$, NH$_3$, HCN, C$_2$H$_2$)} & 6.90 & 4.1\\
     CH$_4$ \& NH$_3$ \tiny{(H$_2$O, CO, CO$_2$, HCN, C$_2$H$_2$)} & -0.92 & undefined\\
     HCN \& C$_2$H$_2$ \tiny{(H$_2$O, CH$_4$, CO, CO$_2$, NH$_3$)}  & -0.71 & undefined \\
     H$_2$O \tiny{(CH$_4$, CO, CO$_2$, NH$_3$, HCN, C$_2$H$_2$)}   & 17.10 & 6.2 \\
     Monotonically Decreasing T \tiny{(all gases)}  & 27.10  & 7.7  \\
     \hline
\end{tabular}
\end{table}

\subsection{Impact of TP profile parameterization}\label{sec:TP_par}
Given that one of the controversies surrounding the dayside spectrum of HD~209458b is whether or not it possesses a stratospheric thermal inversion, it is thus prudent to explore how our thermal profile parameterization could influence our conclusion.  In the nominal model, described above, we used an analytic 5 parameter gray radiative equilibrium description for the thermal profile. Here we use very much simpler profile with far fewer assumptions (similar to Schwarz et al. 2015). The atmosphere is divided up into 3 regions: a deep isothermal region defined by a deep temperature parameter, and two ``T linear-in-log(P)" regions defined by a slope and two log-pressure free parameters defining the break between the three regions, for a total of 5 parameters. Continuity between the regions is enforced. This simple parameterization does not abide by any physical rules (e.g., radiative equilibrium) and is strictly guided by the data and parameter priors.  All other aforementioned model assumptions remain the same.  

\begin{figure*}[h]
\centering
\includegraphics[width=1.0\linewidth]{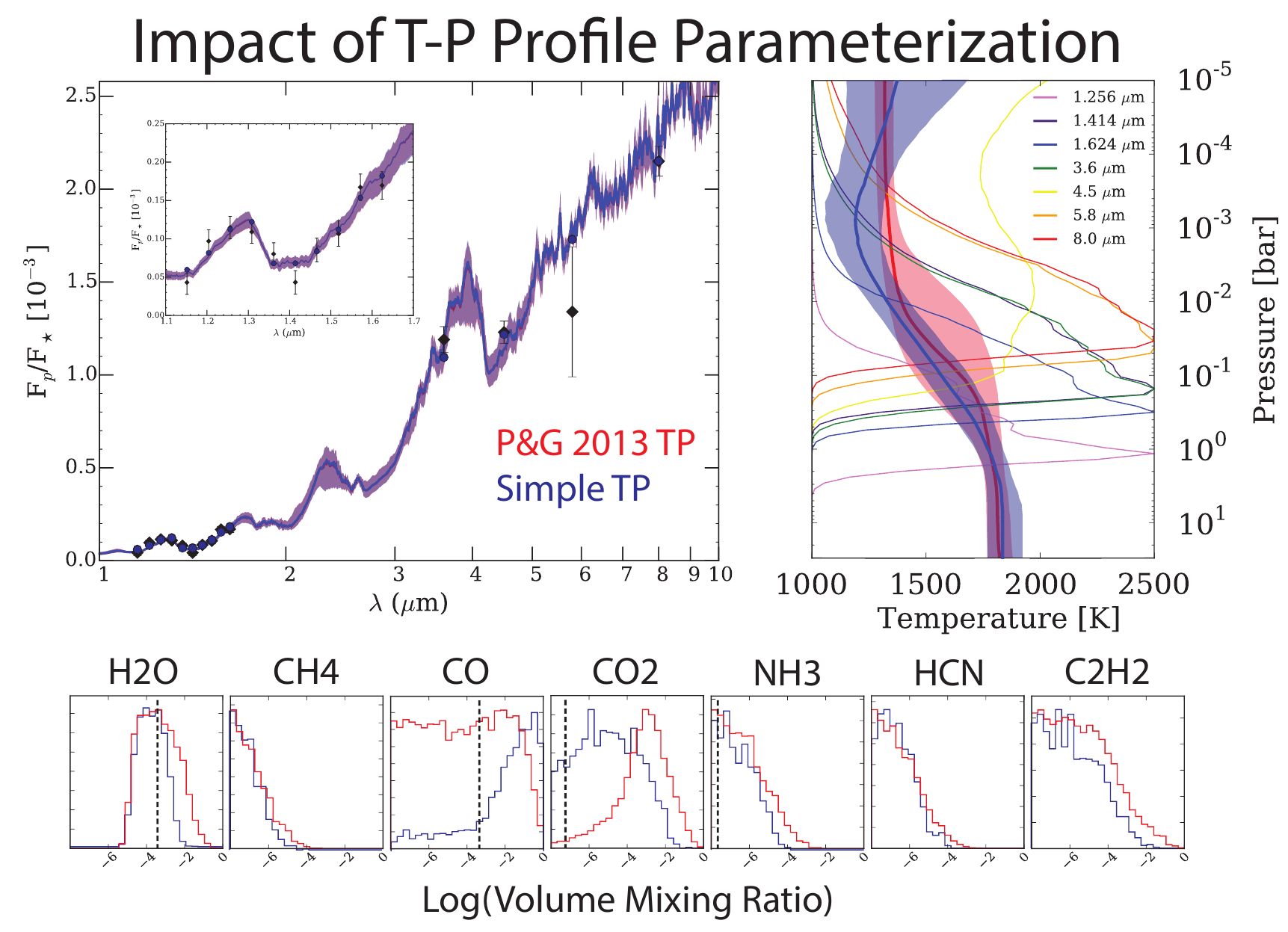}
\caption{\label{fig:spectra_TP_compare} {Impact of the temperature profile parameterization on the retrieval results. Plot structure similar to \ref{fig:spectra_TP}.  In all panels, red corresponds to the nominal Parmentier \& Guillot 2013 (P\&G2013) analytic radiative equilibrium parameterization, and blue to the ``simple" TP parameterization described in \S (Simple).   The spectra resulting from each TP-profile parameterization are nearly indistinguishable but there are noticeable differences in the retrieved TP-profiles and molecular abundances. The temperature gradients over the photosphere are largely consistent, though the uncertainties are smaller within the simple TP parameterization.  Finally, the abundances are largely consistent with the exception of CO \& CO$_2$. See \S\ref{sec:TP_par} for details.
}}
\end{figure*}

Figure \ref{fig:spectra_TP_compare} compares the results from this alternate TP-profile parameterization with our nominal results from \S\ref{sec:results}. In general, the conclusions are unchanged: water is observed to be at or near solar composition, upper limits obtained for CH$_4$,  NH$_3$, HCN, \& C$_2$H$_2$, and there is similar temperature gradient over the photosphere (1 bar - 1 mbar). In fact, upon inspecting the spectra in Figure \ref{fig:spectra_TP_compare} , we find virtually no difference between the fiducial model fits and the simple TP-profile fits. There are a few subtle differences, however: CO is now favored over CO$_2$ (though both are still largely unconstrained), the decreasing TP-profile extends to slightly lower pressures and is better constrained, and the water abundance constraint is 30\% smaller ($\pm$0.7 orders of magnitude). The change in the CO/CO$_2$ abundance distributions is likely due to the slightly steeper temperature gradient over the 4.5 $\mu$m bandpass resulting from the simple TP-profile parameterization, though solar composition abundances are captured well within their 68\% confidence interval.  Again, any inference about these molecules is entirely driven by a single \emph{Spitzer} point, and are thus highly sensitive to any model assumptions. This is because the spectral slopes over the 4.5 $\mu$m bandpass are quite steep. Changing the abundance of both increases the depth of the ``V"-shaped absorption feature, but each has a different effect on the spectral slopes. CO more strongly affects the blue edge of the bandpass where-as CO$_2$ affects the red edge. The temperature gradient effects the slope at both edges. The integrated flux is dependent upon the depth and both the red and blue slopes making for a complicated degeneracy. Any resulting conclusions about the CO/CO$_2$ abundances are therefore not robust.  

\subsection{Impact of Dayside Clouds}\label{sec:cloud}
\begin{figure*}[h]
\centering
\includegraphics[width=1.0\linewidth]{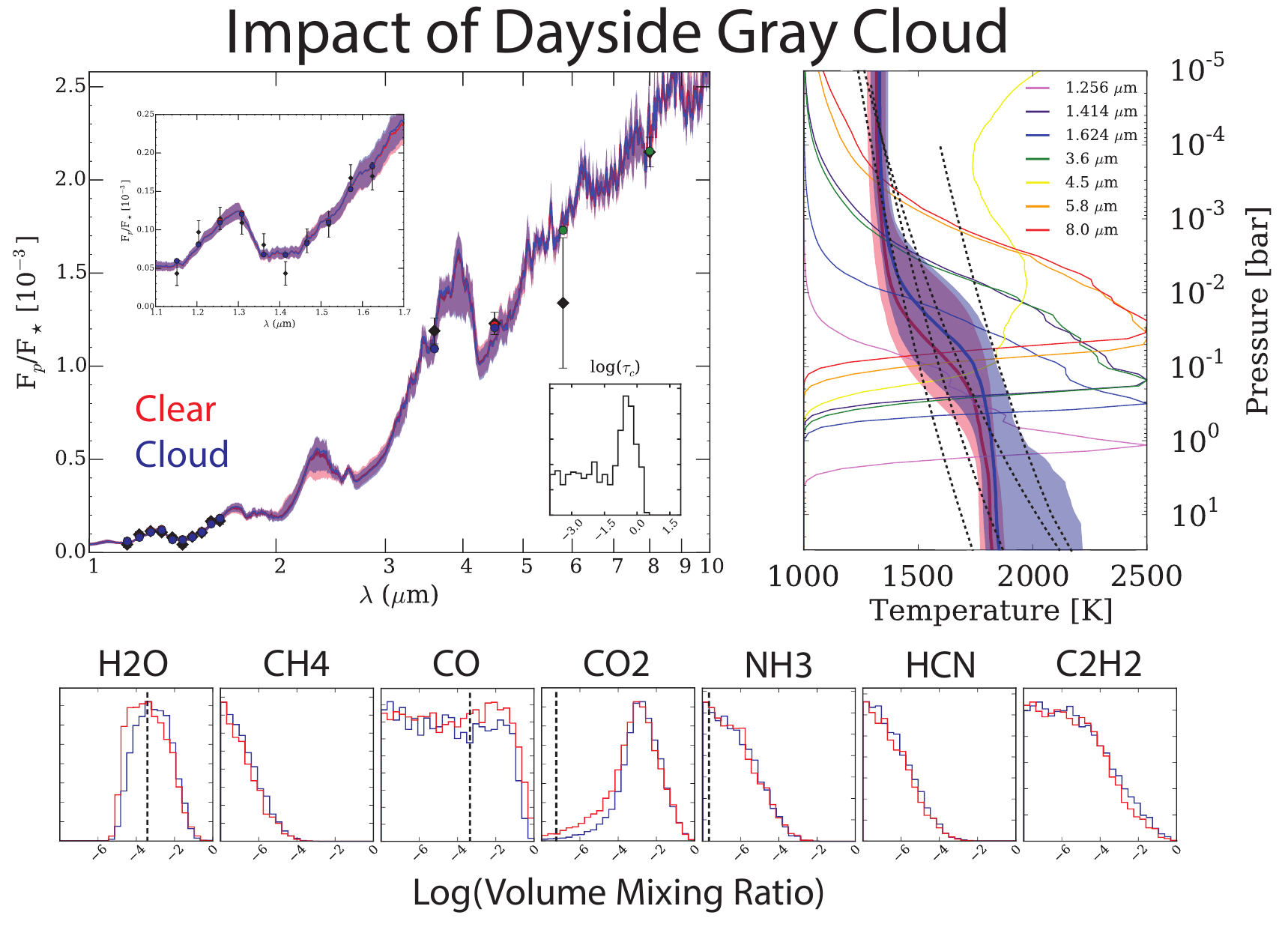}
\caption{\label{fig:spectra_Cloud_compare}{Impact of a dayside cloud. Plot structure similar to \ref{fig:spectra_TP}.  In all panels, red corresponds to the nominal cloud free model (clear) and blue to the model that includes a cloud (cloud) .  The log($\tau_{c}$) inset shows the integrated cloud optical depth over the pressure levels probed by the water band in the WFC3 data.  The integrated optical depths typically fall below unity, indicative of a fairly transparent atmosphere.  Condensate curves for MgSiO$_3$, MgSiO$_4$, Fe, and TiO (from cold to hot) are shown as the dashed lines. While the permitted range of TP-profiles crosses many condensation curves, we find, to high confidence, no evidence for a dayside cloud within the photospheric region. See \S\ref{sec:cloud} for details.
}}
\end{figure*}

Clouds have largely been ignored when interpreting dayside {\it thermal} emission spectra of transiting exoplanets, given that there has not yet been any observational evidence to justify otherwise.  This is surprising given that clouds strongly impact the spectra of similar temperature brown dwarfs, and the mounting evidence for clouds in transmission spectra, and in reflected light curves from Kepler. While clouds can dominate transmission spectra due to the enhanced path length through the atmosphere (Fortney 2005), dayside inhomogeneities (or patchiness) driven by circulation (Parmentier et al. 2016), could weaken the effect of clouds if they are present.     The appearance of the strong water absorption feature over the WFC3 bandpass suggests that the impact of clouds on dayside emission spectrum of HD209458b is minimal. However, to be thorough, we test this hypothesis by introducing a simple, gray, non-scattering cloud in emission with a mass opacity, $\kappa_{c}$, given by
\begin{equation}
\kappa_{c}(P)=\kappa_{P_0}\left(\frac{P}{P_0}\right)^{\alpha}
\end{equation}
where $P_{0}$ is the cloud base pressure, $\kappa_{P_0}$ is the specific absorption coefficient (area/mass) at the base of the cloud, and $\alpha$ is the cloud fill factor index (Burrows, Sudarsky, \& Hubeny 2006).  Figure \ref{fig:spectra_Cloud_compare} summarizes the results.  As anticipated, we find that the cloud has negligible impact on the retrieved abundances and thermal structure, though the retrieved cloudy TP-profile is ever so slightly shifted ``upwards" in pressure, and has greater uncertainty in the deep atmosphere. The Bayes factor between the cloudy model and clear model (0.15--extremely insignificant), suggests that a cloud is not required or justified. The histogram inset in Figure \ref{fig:spectra_Cloud_compare} shows the integrated cloud optical depth \footnote{The photosphere for each sample occurs at a different pressure level. Therefore to compute the integrated cloud optical depth we convolved the cloud optical depth {\it profiles} with the normalized 1.256 (continuum) and 1.414 (water band) $\mu$m contribution functions. This is a rough approximation to the integrated optical depth over the pressure levels probed by the WFC3 observations.  } over the pressure levels probed by the WFC3 water band for 1000 randomly drawn samples. Most of the optical depths are below unity, suggesting small to negligible cloud opacities, again consistent with our Bayes factor results.  Though no impact/evidence for dayside clouds are found here, this may not always be the case, thus a more thorough examination of the emission spectra of multiple planets is necessary to determine the impact of dayside clouds.

\subsection{Impact of \emph{Spitzer} Data Reduction Technique}\label{sec:spitzer}
\begin{figure*}[h]
\centering
\includegraphics[width=1.0\linewidth]{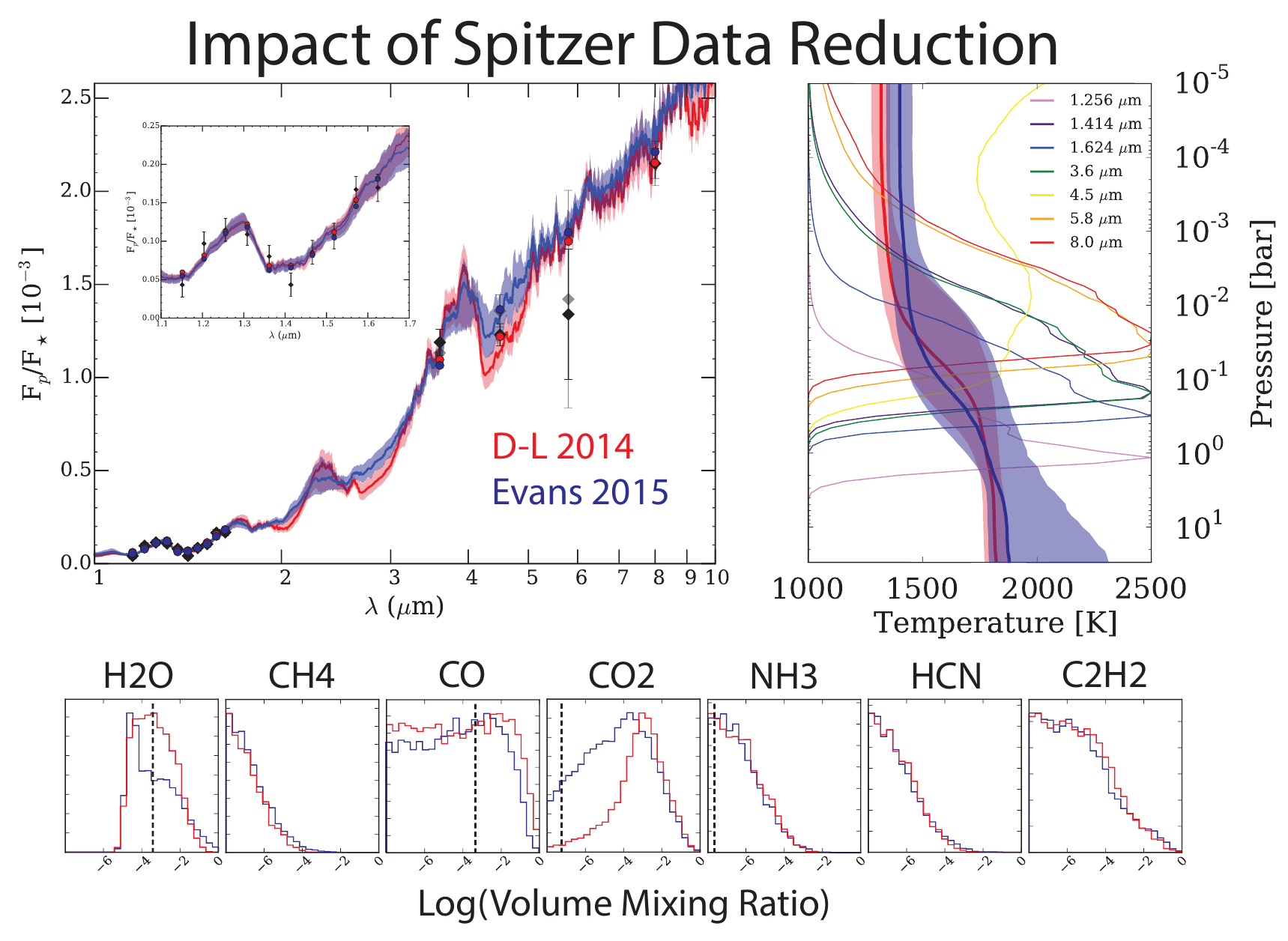}
\caption{\label{fig:spectra_Spitzer_compare}{Impact of choice of \emph{Spitzer} photometry on the retrieval results. Plot structure similar to Figure \ref{fig:spectra_TP}.  In all panels, red corresponds to the nominal model using the Diamond-Lowe et al. (2014) data (thick black points, \S\ref{sec:results}) and blue corresponds to the results derived when using the Evans et al. (2015) data (gray points).   Note the shallower ``V" shaped feature near 4.5 $\mu$ resulting from the increased F$_{p}$/F$_{star}$ in the Evans et al. (2015) data (blue) .  The Evans et al. (2015) data results in somewhat larger temperature uncertainty in the deeper atmosphere and a slightly warmer upper atmosphere.  Overall the molecular abundances are consistent with the except of a widening of the CO$_2$ histogram, owing to the larger 4.5$\mu$m uncertainty on the Evans et al. (2015) datum, and slightly higher deeper eclipse depth.
}}
\end{figure*}

The robustness of published \emph{Spitzer} results have been called into question on multiple occasions (Hansen et al. 2014 (and comparison of references there-in), Diamond-Lowe et al. 2014, Lanotte et al. 2014) and photometric flux values for many planets have changed by several sigma.  As such, we feel it is necessary to explore how the differences in published dayside photometry impacts our results.  For this we compare the retrieved quantities when using the \emph{Spitzer} photometry presented in Diamond-Lowe et al. (2014) (and what we use as our nominal) to those resulting from the photometry published in Evans et al. (2015) (Figure \ref{fig:spectra_Spitzer_compare}).   The measured eclipse depths from Diamond-Lowe et al. (2014) and Evans et al. (2015) are consistent (to within 1$\sigma$), but the latter report larger uncertainties in all channels.  This is typical of Gaussian Processes (GP, used in Evans et al. 2015) relative to other methods, as determined in a recent comparison of \emph{Spitzer} data reduction techniques (Ingalls et al. 2016).  GP also ranked lower in repeatability and reliability compared to BLISS mapping (Stevenson et al. 2012), which was the systematic correction technique used by Diamond-Lowe et al. (2014).  For these reasons, we adopt the eclipse depths reported by Diamond-Lowe et al. (2014) to derive our nominal results. 

 The subtle differences in the two reductions is reflected in the fits as the amplitude of the CO/CO$_2$ absorption feature at 4.5 $\mu$m is shallower with the Evans et al. (2015) data.  These differences in general have a small effect on the retrieved quantities. The largest noticeable difference, however, is in the shape of the CO$_2$ histogram which has ``filled" out more at low abundances. The detailed shape of the water histogram changed slightly, with the median shifting by 5\% and the 68\% confidence interval widening by 20\% (in log(H$_2$O)).   This increased water uncertainty is due to the increased temperature uncertainty in the TP-profile around $\sim$100 mbar where the in-water WFC3 contribution functions partially overlap with the 4.5 $\mu$m \emph{Spitzer} contribution function.  We also experimented with the full-phase derived 4.5$\mu$m eclipse depth from Zellem et al. (2014). The eclipse depth falls between the Diamond-Lowe et al. (2014) and Evans et al. (2015) measurements, and thus results in retrieved values (not shown) that fall in between those derived from the Diamond-Lowe et al. (2014) and Evans et al. (2015) data.  This exercise illustrates the complex sensitivity of the retrieved quantities to the observations, especially when the constraint on a given species (e.g., CO, CO$_2$) is entirely dependent upon a single photometric datum. Nevertheless, the main conclusions of this paper (water abundance consistent with solar and no thermal inversion) are unchanged no matter which \emph{Spitzer} data are used.

\subsection{Elemental Abundance Analysis: Chemical Retrieval on Retrieval}\label{sec:ronr}
In order to more rigorously determine the derived {\it elemental} abundances from the molecular abundances (parameterized through the metallicity, [M/H], and carbon-to-oxygen ratio (C/O)) we introduce a new analysis method, {\it chemical retrieval-on-retrieval}.  The {\em chemical retrieval-on-retrieval} approach straddles the line between the classic ``fully data driven"  (Madhusudhan \& Seager 2009; Lee et al. 2012; Line et al. 2012; Barstow et al. 2012; Benneke \& Seager 2013; Line et al. 2014) and ``a priori imposition of physical processes" (Knutson et al. 2014,  Fraine et al. 2014, Kreidberg et al. 2015, \& Benneke (2015) ). This allows us both to obtain the full range of possible molecular combinations that can explain the data, and {\it a posteriori} address what regions of that parameter space are plausible. We note, that this is the approach taken within the solar system community: molecular abundances (or their profiles) are directly retrieved from the data (whether in situ or remotely) and then compared to chemical models (e.g., Moses et al. 2005; Greathouse et al. 2005; Visscher,  Moses, \& Saslow 2010; Orton et al. 2014).

The {\em chemical retrieval-on-retrieval} is a two step process (as its name implies). First we retrieve the molecular abundances and TP-profile directly from the spectroscopic/photometric data making no assumptions about the relationship between the molecular abundances (\S\ref{sec:results}). We then treat the retrieved {\it marginalized} mixing ratio posterior distributions (from \S\ref{sec:results}) as independent data points to which we fit a thermochemical equilibrium model (again using {\tt pymultinest}). From a statistical perspective, this is very much how one treats wavelength dependent eclipse depths as independent data points when performing a classic atmospheric retrieval. The thermochemical model (NASA CEA2, Gordon \& McBride 1994; Moses et al. 2011; Line et al. 2011; Kreidberg et al. 2015; Molli{\`e}re et al. 2015) computes the molecular mixing ratios along a given pressure-temperature profile given [M/H] and C/O, which are the retrieved parameters. 

We consider all species ($\sim$2000) in the CEA2 thermodynamic library that contain H, He, C, N, O, S, P, Fe, Ti, V, Na, K and assume pure equilibrium (no rainout).  Rather than assume Gaussian distributions for the mixing ratio data (e.g., as one does when using chi-squared), we evaluate the probability density of each thermochemically-computed mixing ratio with the normalized histograms themselves, through interpolation. When computing the net log-likelihood for multiple gases, we simply sum the log of their probability densities together \footnote{This process is the equivalent of computing chi-square for normal distributions}.  We propagate the uncertainty in the retrieved TP-profile from the atmospheric retrieval into the retrieved metallicity and C/O by drawing a random TP-profile from the TP-posterior at each {\tt pymultinest} likelihood evaluation.  This ignores the correlations between the {\it elemental} abundances and the TP profile, however this is acceptable as the marginalized molecular abundances contain that information.  We integrate out vertical profile information by column averaging the thermochemically derived mixing ratio profiles over the infrared photosphere (3bars -1 mbar), as low resolution spectra are largely insensitive to vertical profile information (e.g., Lee et al. 2012; Barstow et al. 2013).

\begin{figure*}[h]
\centering
\includegraphics[width=1.0\linewidth]{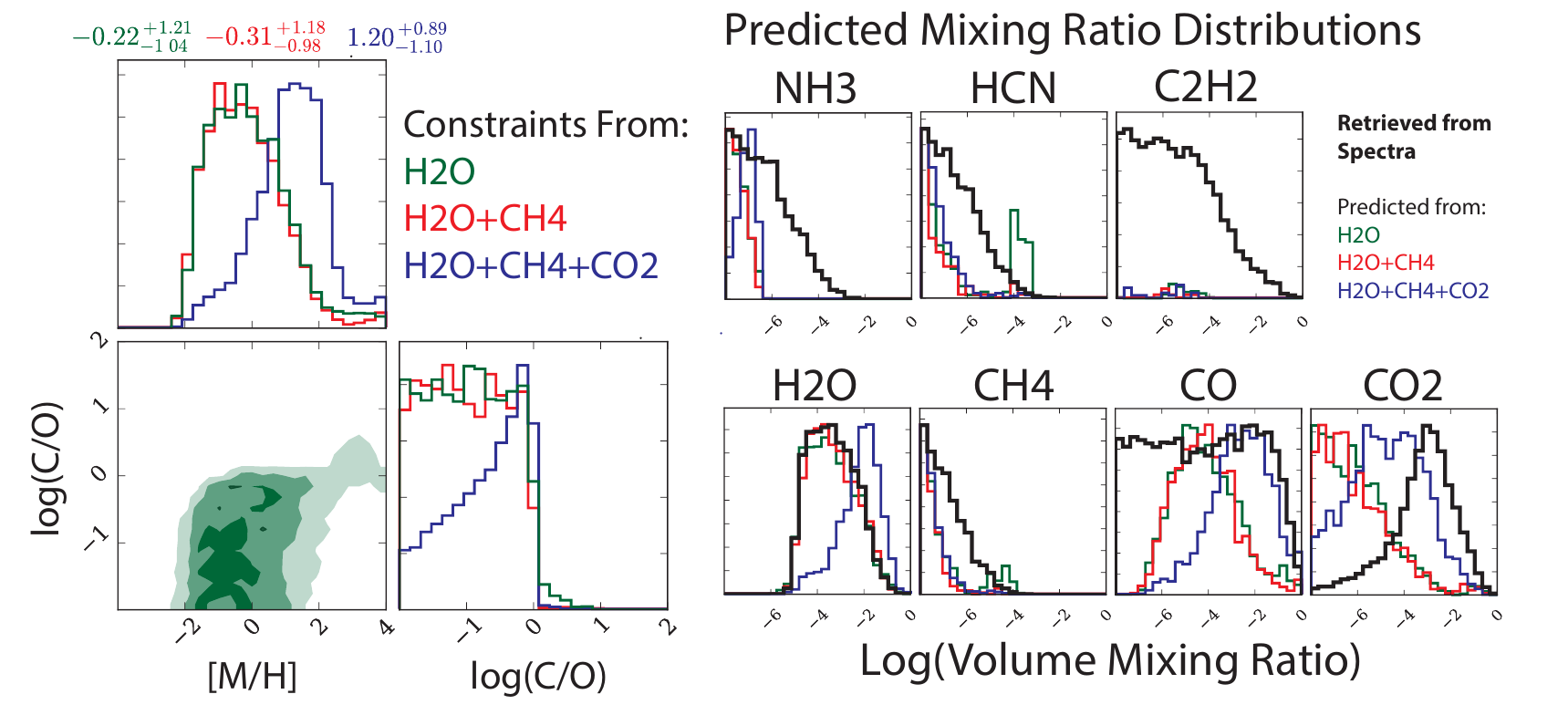}
\caption{\label{fig:RoR}{Results of the {\em chemical retrieval-on-retrieval} method used to derive the atmospheric metallicity ([M/H]) and carbon-to-oxygen ratio (C/O).  [M/H] and log(C/O) are determined using three different combinations of the spectrally retrieved molecular abundances: H$_2$O only (green), H$_2$O+CH$_4$ (red), and H$_2$O+CH$_4$+CO$_2$ (blue).  Left: Pairs plot illustrating the full [M/H]-log(C/O) posterior for the three scenarios. The green 2D histogram shows the correlation between [M/H] and log(C/O) when using only the spectrally retrieved H$_2$O histogram.  Right: Predicted mixing ratio distributions based upon the chemically retrieved [M/H] and log(C/O) when using H$_2$O only (green), H$_2$O+CH$_4$ (red), and H$_2$O+CH$_4$+CO$_2$ (blue) as the data. The thick black histograms are the spectrally retrieved mixing ratio distributions (Figure \ref{fig:spectra_TP}). These results indicate that all spectrally retrieved species, except CO$_2$, are consistent with chemical equilibrium near solar elemental abundances.  The spectrally retrieved CO$_2$ abundances require somewhat higher metallicities, but results in somewhat higher water abundances relative to what is spectrally retrieved.    
}}
\end{figure*}

Within this framework, we investigate how the constraints on each molecular species contribute to our knowledge of the metallicity and C/O. We first ask how well we can constrain the metallicity and C/O from the retrieved water abundance alone, followed by water combined with methane, and finally water combined with methane and carbon dioxide.  Figure \ref{fig:RoR} summarizes the results.    The left shows the metallicity and C/O constraints obtained for each of the three scenarios. The right panel shows the mixing ratios derived from the retrieved metallicity and C/O under thermochemical equilibrium (colored histograms), within the photosphere, compared with the mixing ratios retrieved directly from the data (thick black histograms).  

When using only water to constrain the elemental abundances we find a metallicity range of 0.06 - 9.8 $\times$ solar\footnote{This is a more accurate reflection of the true atmospheric metallicity than simply comparing the retrieved water abundance to what the water abundance should be at solar composition in thermochemical equilibrium, as this method takes into account the degeneracy between the C/O and metallicity on the water abundance. Lower C/O's require a lower metallicity to give the same water abundance, and vice versa}.  There still remains, at 3$\sigma$, a possibility of higher metallicities. We also find, unsurprisingly, that the water histogram derived from the chemical retrieval is nearly identical to that retrieved from the data, that is, the chemical model is providing a ``good fit" to the mixing ratio ``data".  Because there are no carbon species fit for in this scenario, there is a small trickle of probability beyond C/O = 1. In the predicted mixing ratio distributions this low probability high C/O results in a low probability tail of appreciable amounts of HCN and C$_2$H$_2$.     However, when we include the upper limit derived from the methane mixing ratio, the high C/O tail is eliminated, also eliminating the HCN and C$_2$H$_2$ possibilities.  Water and methane together put a firm upper limit on the C/O ratio, but offer little to constrain any lower limit since there is no lower limit to the methane mixing ratio retrieved from the spectrum. The retrieved water and methane distributions are consistent with the predicted distributions when fit together, suggesting that the retrieved water and methane mixing ratios are physically plausible.   

The predicted CO mixing ratios from the water, and water+methane scenarios are well within the spectrally retrieved CO upper limit. However, we do find that that most of the chemically predicted CO$_2$ probability is significantly below the retrieved CO$_2$ distribution.  This suggests that the chemically retrieved metallicity and C/O, while self-consistently able to explain all the other spectrally retrieved molecular species, is unable to adequately explain the CO$_2$ abundance. Finally, when including the retrieved CO$_2$ mixing ratio, with methane and water, we find that a higher metallicity (1.3 - 123 $\times$ solar) is required; unsurprising since CO$_2$ is strongly sensitive to metallicity.  The requirement to fit the relatively high CO$_2$ abundance via high metallicity also pulls the chemically predicted water abundance to higher values, to the point of not being a particularly good fit to the spectrally retrieved water abundance distribution. In short, the spectrally retrieved CO$_2$ abundance and water abundances are not particularly consistent with each other. 

We note that we obtained a similarly high CO$_2$ abundance for our WASP-43b dayside results (Kreidberg et al. 2014) when using the 3.6 and 4.5 $\mu$m \emph{Spitzer} data.  Obviously it is unphysical for CO$_2$ abundances to be larger than the CO abundance for solar metallicities in thermochemical equilibrium (Prinn \& Barshay 1977; Heng \& Lyons 2016 ).  Perhaps the unusually high CO$_2$ abundance is because we are missing some chemical process that enhances CO$_2$ beyond thermochemical equilibrium. This is unlikely, however, as the most likely processes to do so are chemical quenching via vertical mixing, or photochemistry. Both of these processes are unlikely to significantly enhance the CO$_2$ abundance to this level in an atmosphere of these temperatures (Moses et al. 2011, Figure 3; Cooper \& Showman et al. 2006).  Dynamical-chemical effects are also unlikely to significantly impact the abundances (Cooper \& Showman et al. 2006; Agundez et al. 2014 ). We cannot rule out that there could be some additional unknown chemical mechanism that we cannot think of. More likely, however, is that the retrieved CO$_2$ abundance is not robust. This is supported by high degree of sensitivity of the CO$_2$ posterior to our choice TP-profile parameterization or \emph{Spitzer} data, something to consider for future analyses.  We note in an independent analysis (Feng et al. 2016, submitted) with synthetic data of a similar observational setup (WFC3+Spitzer IRAC 3.6 and 4.5 $\mu$m channels), that the retrieved CO$_2$ abundance is highly sensitive to the particular random noise draw of the 4.5 $\mu$m point.  Again, this suggests that spurious abundances derived from a {\it single} broadband measurement should be taken with a grain of salt.The non-linearities of infrared radiative transfer combined with ultra low spectral resolution conspire to produce unphysical solutions.  We are thus more inclined to believe the chemically-retrieved elemental abundances derived via the other molecular species, which are less impacted by our assumptions. 

To summarize, we find that the {\it chemical retrieval-on-retrieval} method produces self-consistent chemical abundances, in agreement with most of our spectrally retrieved abundances at 0.06 - 9.8 $\times$ solar and C/O $<$ 1.  The method is not able to self-consistency reproduce the spectrally retrieved CO$_2$. However, we question the robustness of the spectrally retrieved CO$_2$ abundances which is largely derived from a single broadband measurement and are therefore highly sensitive to our model assumptions. 

\section{Discussion \& Conclusions}\label{sec:conc}
We present the first high precision \emph{HST} WFC3 measurements of the dayside emission of the canonical hot-Jupiter HD~209458b.  These measurements achieve an ultra-high, photon noise limited precision of 15 ppm providing one of the most robust detections of water (6.2$\sigma$) in a dayside emission spectra. The deep water feature, combined with 4 broadband \emph{Spitzer} observations suggest a monotonically decreasing temperature profile at 7.7$\sigma$ confidence, firmly ruling out the presence of a thermal inversion between 1 bar and 1 mbar.  We tested the robustness of our atmospheric inferences by exploring multiple model assumptions such as the impact of the temperature profile parameterization, influence of a gray dayside cloud, and impact of the source of \emph{Spitzer} photometry. We found that the decreasing temperature profile and water abundance were largely invariant to these assumptions, but that the retrieved CO \& CO$_2$ abundances were strongly assumption dependent, owing to their dependence on a single broadband \emph{Spitzer} measurement (4.5$\mu$m).

\begin{figure}[h]
\centering
\includegraphics[width=1.0\linewidth]{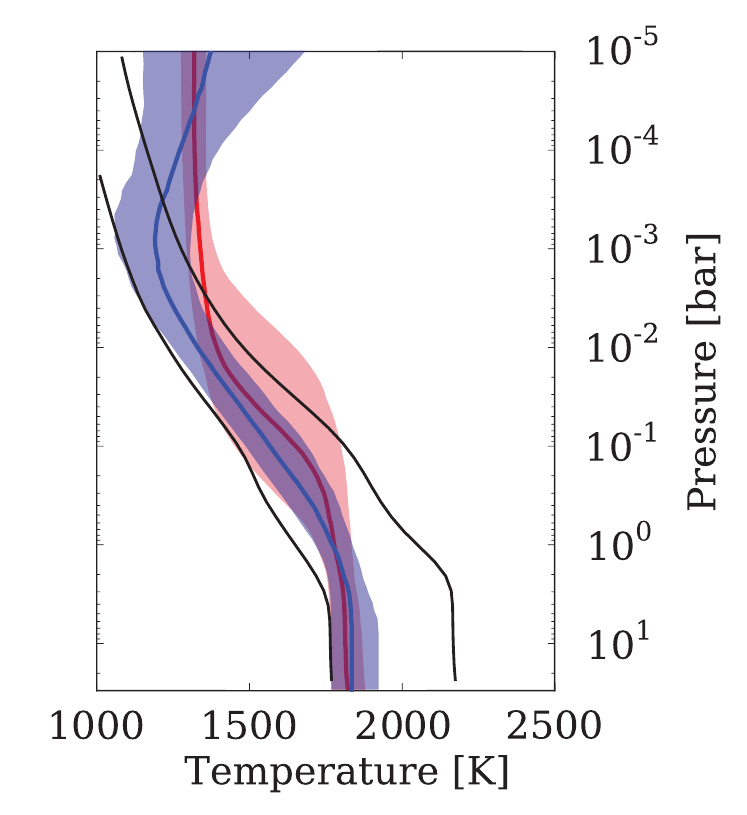}
\caption{\label{fig:compare_TP_fortney}{Comparison of retrieved temperature profiles (under both parameterizations, red and blue) with self-consistent radiative-convective-thermochemical equilibrium models (black curves, Fortney et al. 2008). The cooler self-consistent TP profile corresponds to full redistribution where-as the hotter corresponds to dayside only re-radiation.  The retrieved profiles fall between the two suggesting a redistribution that falls in between.  Overall, the agreement is fairly good in terms of the temperature gradient over the regions probed by the observations (1 bar - 1 mbar).
}}
\end{figure}

In Figure \ref{fig:compare_TP_fortney} we compare the two versions of the retrieved TP profile to two profiles from a self-consistent radiative equilibrium model (Fortney et al. 2005, 2008).  The hotter black profile assumes only redistribution of energy over the day side. The cooler profile is a planet-wide average.  These self-consistent models include equilibrium chemistry and solar abundances, and yield Bond albedos of 0.08.  The retrievals yield a profile intermediate between the radiative equilibrium cases, suggesting either a larger Bond albedo for the planet, or modest energy redistribution to the night side.  The very low geometric albedo from MOST for this planet (Rowe et al. 2008) suggests a low Bond albedo, which favors the explanation that energy is being advected to the night side, as seen in the 4.5 $\mu$m light curve from Zellem et al. (2014).

This relatively ``cool" day side profile is in contrast to our findings for WASP-43b from Stevenson et al. (2014, Figure S5). For that planet we found a profile consistent with little energy loss to the night side, which was in agreement with the low night side flux seen in the full orbit phase curve.  The two planets have nearly equal equilibrium temperatures, but other significant differences: a 4.3 times longer rotation period for HD 209458b (assuming tidal locking), and at 4.8 times larger surface gravity for WASP-43b.  These differences could be important in understanding the the energy balance and redistribution on the two planets.

As discussed in \S\ref{sec:intro}, the claimed detection of a thermal inversion relatively deep in HD~209458b's infrared photosphere launched an active area of research to look for thermal inversions in other planets, search for the shortwave absorber responsible for causing them, and elucidate the phenomenon theoretically. We had previously suggested that the thermal inversion hypothesis for this prototype was not correct based on a re-evaluation of the \emph{Spitzer} photometry that the original inference was based on. Now we show definitively that HD\,209458b does not have a thermal inversion in its infrared photosphere using high precision spectroscopy. The existence of similar thermal inversions in other conventional hot Jupiters (i.e., those with similar temperatures as HD~209458b, say 800 $<$ T$_{eq}$ $<$ 2000\,K), which all have lower quality data, should now be questioned. The recent possible detection of a thermal inversion for WASP-33b (Haynes et al. 2015) is not inconsistent with this statement as this planet exhibits an average dayside brightness temperature of 3000\,K, which means it is in a very different physical and chemical regime as more typical hot Jupiters like HD~209458b.

In retrospect, the thermal inversions proposed for hot Jupiters were always difficult to understand. Perhaps the most serious issue for the thermal inversion hypothesis is that the \emph{Spitzer} photometric bands typically used to discriminate for inversions are expected to probe similar pressure levels for solar-like composition atmospheres in chemical equilibrium (Burrows et al. 2007, 2008; Showman et al. 2009). Therefore, the altitude and absorption strength of the shortwave observer had to be carefully tuned to give the large temperature difference over a small region of the atmosphere that was needed to match the data. The original leading candidate chemical species (TiO and VO) are now thought to not be present in the dayside atmospheres of conventional hot Jupiters because of the strong vertical mixing to keep these heavy molecules aloft at low pressures (Spiegel et al. 2009) and/or the rainout of larger condensate droplets on the cooler nightside (Parmentier et al. 2013) Furthermore, forced inclusion of these species in GCM's could not reproduce the data anyway (Showman et al. 2009).

\begin{figure}[h]
\centering
\includegraphics[width=1.0\linewidth]{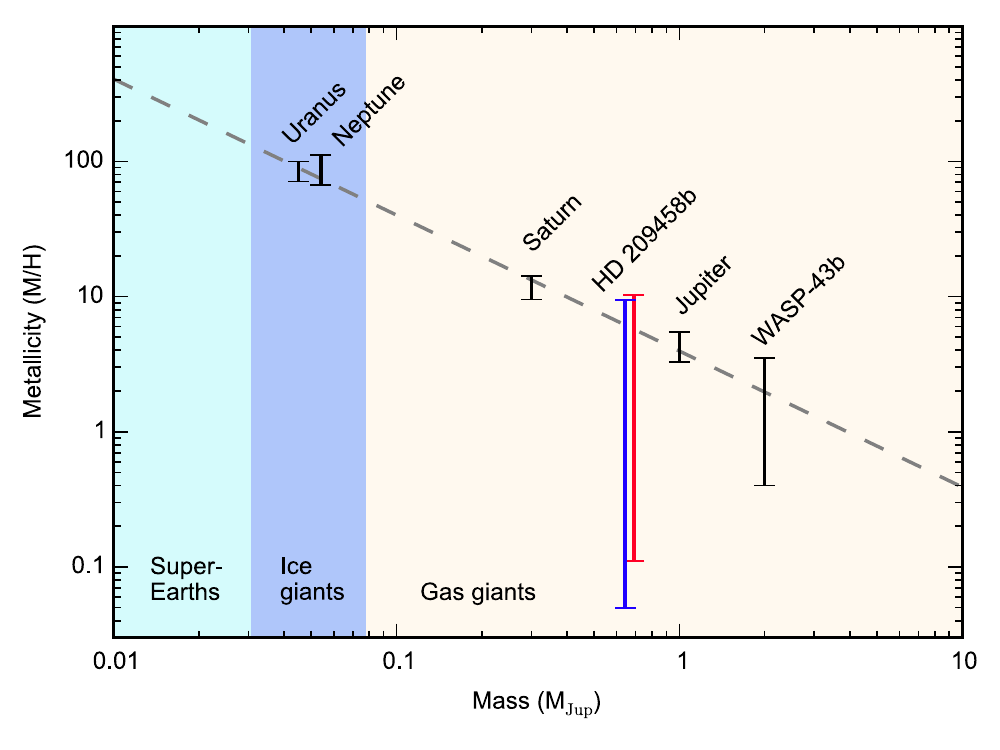}
\caption{\label{fig:mass_metallicity}{Established mass vs. atmospheric metallicity trend seen in the solar system.  We show our previous combined emission+transmisson water derived metallicity result from WASP-43b along with our derived abundances in this work.  For HD~209458b we show two measurements: The red is the water derived metallicity assuming solar carbon-to-oxygen ratios where as the blue is the metallicity derived from the {\it chemical retrieval on retrieval} taking into account the metallicity-C/O degeneracy,  based on the retrieved water abundance. These measurements are consistent with the trend. }}
\end{figure}

In addition to studying the thermal structure of HD~209458b, we also introduced a new elemental abundance analysis, {\it chemical retrieval-on-retrieval} where we {\it a posteriori} fit the retrieved molecular abundances with a thermochemical model to determine the atmospheric metallicity and carbon-to-oxygen ratio.  From this analysis we find that the measured water abundance from our spectrum of HD~209458b is consistent with 0.06 - 9.8 $\times$ the solar metallicity, and we firmly rule out C/O values greater than 1.  The metallicity derived from the water abundance measurement (both from the chemical retrieval on retrieval and comparison of the retrieved water abundance to solar) is consistent with the trend of increasing atmospheric metallicity for decreasing planet mass that is observed in our solar system (see Figure \ref{fig:mass_metallicity}). This result extends the agreement that is seen between the solar system trend and exoplanet atmosphere abundances, which was first investigated by Kreidberg et al. (2014a). The C/O upper limit of 1 buttresses the similarly strict limits that were recently obtained for HD~209458b and other hot Jupiters from the interpretation of transit transmission spectra while accounting for clouds (Benneke 2015; Kreidberg et al. 2015).

Our solar water abundance determination for HD~209458b is inconsistent with the strongly sub-solar (0.007 - 0.05 $\times$ solar) abundance interpretation of the planet's WFC3 transmission spectrum (Deming et al. 2013) by Madhusudhan et al. (2014b). However, Madhusudhan et al. (2014b) did not account for clouds in their retrieval. Benneke (2015) and our own unpublished analysis have shown that neglecting clouds in the interpretation of the planet's WFC3 transmission spectrum yields spuriously low and precise abundances because clouds are truncating the water absorption feature compared to the expectation for a clear atmosphere. The water abundance we derive from the dayside emission spectrum of HD~209458b is consistent with, but more precise than the water abundance derived from the transmission spectrum by Benneke (2015). The consistency between water abundance values derived from dayside emission and transmission spectra is expected for hot Jupiters on thermochemical grounds (Moses et al.\ 2011) and was previously seen for the only other planet for which this comparison as been made (WASP-43b, Kreidberg et al. 2014a). We also find that our retrieved molecular abundances are thermochemically self-consistent, and thus ``physical", with the exception of CO$_2$, which is best explained with metallicities $>10\times$ solar. However, again, the retrieved CO$_2$ abundance is strongly model and data dependent, suggesting that robust inference of CO$_2$ is not yet possible.  

Of the 7 planets with published \emph{HST} WFC3 {\it emission} spectra (WASP-12b (Swain et al. 2013) WASP-4b, TrES-3b (Ranjan et al. 2014), HD~189733b (Crouzet et al. 2014), WASP-43b (Kreidberg et al. 2014b), WASP-33b (Haynes et al. 2015)) only two, WASP-43b, and HD~209458b (this work) report a robust detection of water over the WFC3 bandpass.    Detecting or not detecting water has significant implications for the nature of the thermal structure and/or composition.  Over the WFC3 bandpass alone, it is difficult to determine whether or not the lack of a water absorption feature is due to isothermal atmospheres (or nearly isothermal), an intrinsic depletion of water, either due to high C/O or low metallicity, or an optically thick dayside cloud at high altitudes. This ambiguity emphasizes the need to push for higher precision emission observations to determine the presence, or lack-there-of, of water absorption over a wide range of stellar compositions (C/O, metallicity), and planetary equilibrium temperatures. Furthermore, by combining emission and transmission observations (e.g., Kreidberg et al. 2014a) we can begin to break the temperature profile - water - cloud degeneracy.  


\section{Acknowledgements}
These observations were made under the GO Treasury Program 13467 with the NASA/ESA Hubble Space Telescope at the Space Telescope Science Institute which is operated by the Association of Universities for Research in Astronomy, Inc., for NASA, under the contract NAS 5-26555. M.R.L. acknowledges support provided by NASA through Hubble Fellowship grant 51362 awarded by the Space Telescope Science Institute.  K.B.S Acknowledges support from the NASA Exoplanet Science Institute Sagan Postdoctoral Fellowship. J.L.B. acknowledges support from the David and Lucile Packard Foundation. We also thank Dan Foreman-Mackey for making the {\it corner.py} plotting routine available to the public. We thank Drake Deming, Peter McCullough, Adam Burrows, Sara Seager, David Charbonneau, and Derek Homeier for being co-investigators on the \emph{HST} observing proposal.

\end{document}